\documentclass[aps, prd, twocolumn, showpacs, preprintnumbers, letterpaper, amsmath, amssymb, superscriptaddress, nofootinbib, 10pt]{revtex4-1}

\usepackage{graphicx}
\usepackage{pgfplots}
\usepackage{epsfig}
\usepackage[latin1]{inputenc}
\usepackage{color}
\usepackage{changes}
\usepackage{amsmath,amsfonts,amssymb,dcolumn,bm,hyperref,xspace}
\usepackage{multirow}
\usepackage{threeparttable}
\usepackage{graphicx}

\begin{document}
\title{Cosmological constant constraints from observation-derived energy condition bounds and their application to bimetric massive gravity}

\author{M.E.S. Alves}
\email{marcio.alves@ict.unesp.br}
\affiliation{Universidade Estadual Paulista (UNESP), Instituto de Ci\^encia e Tecnologia, S\~ao Jos\'e dos Campos, SP, 12247-004, Brazil}

\author{F.C. Carvalho}
\email{fabiocabral@uern.br}
\affiliation{Instituto Nacional de Pesquisas Espaciais, Divis\~ao de Astrof\'\i sica, Av. dos Astronautas 1758, S\~ao Jos\'e dos Campos, SP 12227-010, Brazil}
\affiliation{Universidade do Estado do Rio Grande do Norte, Mossor\'o, 59610-210, RN, Brazil}

\author{J.C.N. de Araujo}
\email{jcarlos.dearaujo@inpe.br}
\affiliation{Instituto Nacional de Pesquisas Espaciais, Divis\~ao de Astrof\'\i sica, Av. dos Astronautas 1758, S\~ao Jos\'e dos Campos, SP 12227-010, Brazil}

\author{M. Penna-Lima}
\email{penna@lapp.in2p3.fr}
\affiliation{Laboratoire d'Annecy de Physique des Particules (LAPP), Universit\'e Savoie Mont Blanc, CNRS/IN2P3, F-74941 Annecy, France}
\affiliation{Universidade de Bras\'ilia, Instituto de F\'isica, Caixa Postal 04455, Bras\'ilia, DF, 70919-970, Brazil}
\affiliation{Centro Brasileiro de Pesquisas F\'isicas,  Rua Dr. Xavier Sigaud 150, Rio de Janeiro, 22290-180, RJ, Brazil}

\author{S.D.P. Vitenti}
\email{sandro.vitenti@uclouvain.be}
\affiliation{Centre for Cosmology, Particle Physics and Phenomenology, Institute of Mathematics and Physics, Louvain University, 2 Chemin du Cyclotron, 1348 Louvain-la-Neuve, Belgium}
\affiliation{Institut d'Astrophysique de Paris, GReCO, UMR7095 CNRS, 98 bis boulevard Arago, 75014 Paris, France}

\date{\today}

\begin{abstract}
Among the various possibilities to probe the theory behind the recent accelerated expansion of the universe, the energy conditions (ECs) are of particular interest, since it is possible to confront and constrain the many models, including different theories of gravity, with observational data. In this context, we use the ECs to probe any alternative theory whose extra term acts as a cosmological constant. For this purpose, we apply a model-independent approach to reconstruct the recent expansion of the universe. Using Type Ia supernova, baryon acoustic oscillations and cosmic-chronometer data, we perform a Markov Chain Monte Carlo analysis to put constraints on the effective cosmological constant $\Omega^0_{\rm eff}$. By imposing that the cosmological constant is the only component that possibly violates the ECs, we derive lower and upper bounds for its value. For instance, we obtain that $0.59 < \Omega^0_{\rm eff} < 0.91$ and $0.40 < \Omega^0_{\rm eff} < 0.93$ within, respectively, $1\sigma$ and $3\sigma$ confidence levels. In addition, about 30\% of the posterior distribution is incompatible with a cosmological constant, showing that this method can potentially rule it out as a mechanism for the accelerated expansion. We also study the consequence of these constraints for two particular formulations of the bimetric massive gravity. Namely, we consider the Visser's theory and the Hassan and Roses's massive gravity by choosing a background metric such that both theories mimic General Relativity with a cosmological constant. Using the $\Omega^0_{\rm eff}$ observational bounds along with the upper bounds on the graviton mass we obtain constraints on the parameter spaces of both theories.	
\end{abstract}

\maketitle

\section{Introduction}
\label{sec:introd}

The currently observed acceleration of the Universe is inferred, for example, from the measurements of luminosity distance as a function of redshift for distant Type Ia supernovae (SNe Ia) \cite{Riess1998, Perlmutter1999}. This result has been supported by other cosmological observations such as the anisotropies in the Cosmic Microwave Background radiation (CMB) \cite{Hinshaw2013, PlanckCollaboration2015}, the Baryon Acoustic Oscillation (BAO) \cite{Eisenstein2005} and the Large Scale Structure (LSS) \cite{Tegmark2006,DES2017} data. Two possible approaches to describe this phenomenon consist in modifying the theory of gravitation or to include new fields in the matter-energy content of the universe (see \cite{Frieman2008} and references therein). Following the former, an attractive method to tackle this problem is to consider possible modifications of the Einstein's theory of general relativity (GR), considering for example that gravitons are massive particles (see, e.g., \cite{Fierz1939, Visser1998, Rubakov2008, Hinterbichler2012}).

Independently of the method used to approach the acceleration phenomenon, in most cases, when dealing with homogeneous and isotropic background, one can rewrite the equation of motion for the scale factor as a Friedmann equation with an extra term. For this reason it is useful to study this extra term phenomenologically. Then, once its properties are constrained by the data, one can study their consequences for specific theories behind the aforementioned term. In this work we follow this approach focusing on the consequences of assuming that the extra term behaves as a cosmological constant. Subsequently, we apply the obtained constraints for two distinct formulations of massive bimetric gravity.

The massive graviton problem has been studied since the seminal work by Fierz and Pauli (FP) in 1939, when they first wrote the action for a Lorentz invariant massive spin-2 theory~\cite{Fierz1939}. The linear FP model coupled to a source was studied by \citet{Dam1970} and \citet{Zakharov1970, Zakharov1970a} (vDVZ), who discovered the surprising fact that the FP model differs from GR even when the graviton is massless. This problem, known as vDVZ discontinuity, can be avoided by the Vainshtein mechanism, which takes into account the non-linearities of the FP model~\cite{Vainshtein1972}. However, soon after the Vainshtein findings, Boulware and Deser (BD) studied some specific fully nonlinear massive gravity theories and showed that they possess ghost-like instabilities~\cite{Boulware1972}. Since then, the problem of the ghost-like solutions in massive gravity theories have been extensively debated in the literature \citep[see,e.g.,][]{Nieuwenhuizen1973,Rham2011,Hassan2012a,Hassan2012}. See also the review \cite{Rham2014} and references therein.

Consequently, the extension of massive gravity to strong fields (non-linear order) is not a trivial task. One attempt was carried out by M. Visser by introducing a background metric that is not subjected to any dynamical equation \cite{Visser1998}. In his theory the mass term depends both on the dynamical and background metrics such that, in the linear limit, the massive field obeys a Klein-Gordon equation with a source term, and the full GR is recovered when the graviton mass vanishes. However, in Ref.~\cite{Roany2011}, de Roany {\it et al.} have pointed out that the Visser massive graviton tensor must be corrected by a factor equal to the square root of the ratio of the determinants of the background and of the dynamical metrics. They claimed that a multiplicative factor is missing in the graviton tensor originally derived by Visser, which has no consequences on the weak field approach but becomes important in the strong field regime when, for instance, cosmological applications are considered.  

Nevertheless, it is well-known that bimetric gravity theories are generally affected by the same ghost instability appearing in massive gravity \cite{Baccetti2013}. Recently, however, the existence of a consistent ghost-free bimetric theory of massive gravity was demonstrated by \citet{Hassan2012}. It was shown that the theory is ghost-free at the complete non-linear level \citep[see also][]{Rham2010,Rham2011,Koyama2011,DAmico2011,Hassan2012a,Kugo2014,Rham2014a}. The remaining presence of a BD ghost for a family of theories, due to the vanishing Hessian of their actions, was discussed in Ref.~\cite{Chamseddine2013}. Notwithstanding, Alexandrov confirmed that those theories are ghost-free~\cite{Alexandrov2014}.

An analysis of the relationship between massive gravity and bimetric gravity (the so called bigravity) in the context of the \citet{Hassan2012} approach was carried out by \citet{Baccetti2013}, focusing on a procedure which allows massive gravity to be treated as a suitable limit of bigravity. In essence, we can say that in the limit of a vanishing kinetic term for the background metric, the solutions of bigravity will also be solutions of massive gravity compatible with a non-flat-background metric, whereas the opposite is not necessarily true.

Recently, different cosmological applications of bigravity and massive gravity models have been analysed \citep[see e.g., ][]{Comelli2012, Comelli2012a, Volkov2012, Strauss2012, Koennig2014, Solomon2015, Akrami2015}. Particularly, \citet{Strauss2012} have considered cosmological solutions of bigravity that reproduce the current cosmic acceleration and fitted such models to observational data like SNIa, CMB and BAO. \citet{Koennig2014} constrained the parameters of bigravity using SNeIa data, and they found out a number of simple rules for viable cosmological models that lead to a final de Sitter cosmological state. The cosmological viability of bigravity has also been examined by \citet{Akrami2015}. Exploring a region of the parameter space, overlooked so far, they showed that the model provides late-time acceleration in agreement with observations. 

In Refs.~\cite{Alves2008, Alves2010a, Alves2011, Basilakos2011} the authors have shown that Visser's massive gravity could be a viable explanation of the late-time acceleration phase of the Universe. These work have demonstrated that the predicted growth rate of clustering as well as the shape and amplitude of the redshift distribution of cluster-size halos are slightly different from those obtained in the $\Lambda$CDM cosmology. Therefore, these different signatures could be compared and tested against observations. 

One way to probe a theory of gravitation is to compute its respective energy conditions (ECs) and confront them with the observational data. In the context of GR the ECs bounds were scrutinized using SNeIa data \cite{Santos2007, Lima2008, Lima2008a}. A number of authors also studied the ECs in alternative theories of gravitation. For instance, the ECs have been used to constrain $f(R)$ theories of gravity~\cite{Santos2007b, Reboucas2009, Atazadeh2014}, and extensions involving nonminimal curvature couplings~\cite{Bertolami2009, Wang2010, Garcia2010, Garcia2011, Wang2012, Wu2014}. Bounds on modified Gauss-Bonnet $f(G)$ gravity from the ECs have also been analyzed~\cite{Wu2010a, Garcia2011a, Garcia2011b}. The recently proposed $f(R,T)$ theories of gravity has been considered~\cite{Harko2011,Alvarenga2013, Sharif2013}. The bigravity theory has also been studied in the same context~\cite{Baccetti2012} and a possible violation of the null EC was found.

In this work, we present the ECs and use them to constrain, in the cosmological scenario, the extra effective term (working as a cosmological constant), and also the bimetric massive gravity theories that come from the recent approaches by \citet{Roany2011} (Visser's Lagrangian) and \citet{Baccetti2013} (Hassan and Rosen's Lagrangian). 
Here we introduce a different idea, instead of simply testing the ECs, we assume that the remaining matter content (baryons, photons, neutrinos, cold dark matter, etc) fulfills the ECs. By doing this, we are now able to obtain lower and upper bounds for the cosmological constant term.
Then, we confront these bounds with estimates of the deceleration and Hubble functions, which are reconstructed as functions of the redshift by using the model-independent approach presented in Ref.~\cite{Vitenti2015a}. This reconstruction makes use of the Sloan Digital Sky Survey-II and Supernova Legacy Survey 3 years (SDSS-II/SNLS3) combined with the Joint Light-curve Analysis SNe Ia sample (JLA)~\cite{Betoule2014}, BAO data~\cite{Beutler2011, Font-Ribera2014, Ross2015, Delubac2015, Alam2016, Ata2017} and $H(z)$ measurements~\cite{Stern2010, Riess2011, Riess2011a, Moresco2012, Moresco2015}. As a result, we are able to find out the admissible regions of the parameter space of both bimetric massive theories in order to fulfill the strong and dominant ECs.
 
The paper is organized as follows: in Sec.~\ref{sec:BMC} we briefly introduce the bimetric massive gravity by using both the Visser's and the Hassan and Rosen's Lagrangian. In Sec.~\ref{sec:EC} we present the ECs for gravitational theories where the extra term acts as a cosmological constant (obtained in details in a general context for a class of extended theories of gravity in the companion paper~\cite{Penna-Lima2017}), and then we derive the EC inequalities for these two bimetric massive gravity theories. In Sec.~\ref{sec:results}, by using the EC bounds estimated from SNe Ia, BAO and $H(z)$ data,  we discuss the constraints imposed on the parameters of these massive gravity theories. Finally, in Sec.~\ref{sec:conclusions} we present our concluding remarks. Throughout the article we use the metric signature $(-,+,+,+)$ and units such that $c = \hbar = 1$ unless otherwise mentioned. 
 
\section{Bimetric Massive Gravity}
\label{sec:BMC}

The graviton mass can be consistently taken into account by imposing the existence of a background metric $f_{\mu\nu}$ in addition to the dynamical metric $g_{\mu\nu}$. In the context of massive gravity the metric $f_{\mu\nu}$ is externally specified and not determined by the theory itself {\cite{Rham2014}. Only the physical metric $g_{\mu\nu}$ couples to matter and determine the geodesics followed by particles. As mentioned in Sec.~\ref{sec:introd}, we consider here two approaches to massive gravity: (i) the one proposed by Visser \cite{Visser1998} and (ii) the massive Lagrangian proposed by Hassan and Rosen \cite{Hassan2012}. These two theories possess some similarities as well as important differences as described below.

In both cases the action can be written in the following form 
\begin{equation}\label{eq:massive_action}
S = \frac{1}{16\pi G}\int d^4x  \left[\sqrt{-g}R(g_{\mu\nu}) + {\cal L}_{mass}(g_{\mu\nu},f_{\mu\nu})\right] + S_{m},
\end{equation}
where $G$ is the gravitational constant, $g$ is the determinant of the dynamical metric $g_{\mu\nu}$, $R(g_{\mu\nu})$ is the Ricci scalar and  $S_{m}$ is the matter action, as usual. Finally, ${\cal L}_{mass}$ is the ``massive'' Lagrangian which is the only quantity depending on both $g_{\mu\nu}$ and $f_{\mu\nu}$. Hence, the mass of the graviton is introduced via an interaction between the dynamical and the background metrics.

In the Visser's approach, the massive Lagrangian is given by \cite{Visser1998}
\begin{align}\label{eq:Visser_Lagrangian}
{\cal L}_{mass} (g_{\mu\nu} &,f_{\mu\nu}) = -\frac{1}{4}m^2 \sqrt{-f} \Big\{ f^{\alpha\beta}f^{\mu\nu}(g_{\alpha\mu} - f_{\alpha\mu}) \nonumber \\
&\times  (g_{\beta\nu} - f_{\beta\nu}) - \frac{1}{2}[f^{\alpha\beta}(g_{\alpha\beta} - f_{\alpha\beta})]^2 \Big\},
\end{align}
where $m$ is the mass of the graviton.
This Lagrangian is essentially motivated by the fact that in the weak field limit, for which $g_{\alpha\beta} = f_{\alpha\beta} + h_{\alpha\beta}$ with $|h_{\alpha\beta}|\ll 1$, we obtain that the field $h_{\alpha\beta}$ obeys the Klein-Gordon equation, when $f_{\mu\nu}$ is taken to be the Minkowski metric $\eta_{\mu\nu}$. In this case, GR is consistently recovered when the graviton mass vanishes.

On the other hand, the bimetric Lagrangian considered by Hassan and Rosen \cite{Hassan2012, Hassan2012a} is a function of the quantity
\begin{equation}
{\gamma^\mu}_\nu = {\left(\sqrt{g^{-1}f}\right)^\mu}_\nu, \quad {\rm i.e.,} \quad {\gamma^\mu}_\sigma {\gamma^\sigma}_\nu = g^{\mu\sigma}f_{\sigma\nu},
\end{equation}
and is given by
\begin{equation}\label{eq:Hassan_Lagrangian}
{\cal L}_{mass} =  2m^2 \sqrt{-g} \left[ e_2(K) - c_3 e_3(K) - c_4e_4(K) \right],
\end{equation}
with ${K^\mu}_\nu = {\delta^\mu}_\nu - {\gamma^\mu}_\nu$. The parameters $c_3$ and $c_4$ are dimensionless, and $e_n(K)$ are elementary symmetric polynomials given by
\begin{align}
e_2(K) &= \frac{1}{2}\left([K]^2 - [K^2]\right), \\
e_3(K) &= \frac{1}{6}\left([K]^3 - 3[K][K^2] + 2[K^3]\right), \\
e_4(K) &= \frac{1}{24}\big([K]^4 - 6[K^2][K]^2 + 3[K^2]^2 \nonumber \\
 &+ 8[K][K^3] - 6[K^4]\big),
\end{align}
where $[K] = {\rm tr}({K^\mu}_\nu)$.

The Lagrangian given by Eq.~\eqref{eq:Hassan_Lagrangian} is the most general ghost free mass term and it is constructed as a ``deformed'' determinant. It can be seen that it is of fourth order in the quantity ${\gamma^\mu}_{\nu}$, and all higher order terms are identically zero in four dimensions (for further discussions, see \cite{Hassan2012a}). Therefore, unlike the Visser's Lagrangian, this theory has two additional parameters, namely $c_3$ and $c_4$, besides the graviton mass $m$.

\section{Energy condition bounds in bimetric massive gravity}
\label{sec:EC}

In order to further study the bimetric massive gravity theories described in Sec.~\ref{sec:BMC}, we now apply the general methodology presented in \cite{Penna-Lima2017}, in the context of extended theories of gravity, to compute the ECs and, then, put constraints on the parameters of these bimetric massive theories using observational data. 

\subsection{Energy conditions}\label{sec IIIA}

As discussed in \cite{Penna-Lima2017} the strong and the null ECs are derived from the Raychaudhuri equation for congruences of timelike and null curves, respectively. That is,
\begin{align}
R_{\mu\nu}t^\mu t^\nu & \geq 0, \label{eq:sec1} \\
R_{\mu\nu}k^\mu k^\nu & \geq 0 \label{eq:nec1},
\end{align}
where $t^\mu$ ($k^\mu$) is a timelike (null) tangent vector field.
Now, considering that only the ordinary matter should obey such conditions, Eq.~\eqref{eq:sec1} can be rewritten in terms of the energy-momentum tensor, namely
\begin{equation}
\left(T_{\mu\nu} - \frac{1}{2}g_{\mu\nu}T\right)t^{\mu}t^{\nu} \geq 0.
\end{equation}

Furthermore, considering the fluid four velocity $U^\nu = (-1,0,0,0)$, the above inequation can be split in the following two inequalities
\begin{equation}\label{eq:SEC_1_for_T}
T_{\mu\nu}U^\mu U^\nu + \frac{1}{2}T \geq 0,
\end{equation}
and
\begin{equation}\label{eq:SEC_2_for_T}
T_{\mu\nu}k^\mu k^\nu \geq 0.
\end{equation}
Note that Eq.~\eqref{eq:nec1} also leads to Eq.~\eqref{eq:SEC_2_for_T}.
Therefore, Eqs.~\eqref{eq:SEC_1_for_T} and \eqref{eq:SEC_2_for_T} express simultaneously the strong energy condition (SEC) and Eq.~\eqref{eq:SEC_2_for_T} expresses the null energy condition (NEC). Thus, the fulfillment of SEC implies also that NEC is fulfilled.

In short, we are assuming that $T_{\mu\nu}$ includes only ordinary matter such as baryons, dark matter, radiation and neutrinos. Consequently, in the absence of modifications, Eqs.~\eqref{eq:SEC_1_for_T} and \eqref{eq:SEC_2_for_T} imply Eqs.~\eqref{eq:sec1} and \eqref{eq:nec1}. This is motivated by the fact that in other scales, where the modification should be irrelevant   (e.g., laboratory, solar system), these matter components fulfill the ECs.

The weak energy condition (WEC) and dominant energy condition (DEC) are restrictions on the energy-momentum tensor $T_{\mu\nu}$. WEC states that
\begin{equation}
T_{\mu\nu}t^\mu t^\nu \geq 0,
\end{equation}
or
\begin{equation}\label{WEC for T}
T_{\mu\nu}U^\mu U^\nu \geq 0, \quad {\rm and} \quad T_{\mu\nu}k^\mu k^\nu \geq 0,
\end{equation}
while DEC,
\begin{equation}\label{DEC for T}
T_{\mu\nu}{T^\nu}_\lambda t^\mu t^\lambda \geq 0,
\end{equation}
states that the speed of the energy flow of matter is less than the speed of light. It is worth noting that DEC includes WEC.

\subsection{Friedmann equations and energy conditions in the bimetric massive gravity} \label{sec iiib}

In Ref \cite{Penna-Lima2017}, we considered a class of extended theories of gravity for which the field equations can be written in the generic form
\begin{equation}\label{eq:field_eqs_1}
G_{\mu\nu} + H_{\mu\nu} = \frac{8\pi G}{g_1} T_{\mu\nu},
\end{equation}
where $G_{\mu\nu}$ is the Einstein tensor with  $T_{\mu\nu}$ being the usual energy-momentum tensor for matter fields. The additional tensor $H_{\mu\nu}$ depends on the details of each theory and can be a function of the metric, of scalar and vector fields and of covariant derivatives of these quantities. The factor $g_1$ drives the modified coupling with the matter fields. Such a class of theories is characterized by the existence of cosmological solutions. In this work we consider the case for which the modified gravity term acts effectively as a cosmological constant and $g_1 = 1$. 

In the context of massive gravity theories the tensor $H_{\mu\nu}$ can be derived in a straightforward manner from the Lagrangians \eqref{eq:Visser_Lagrangian} or \eqref{eq:Hassan_Lagrangian} (we refer the reader to Refs. \cite{Roany2011,Hassan2012a,Baccetti2013}). In order to show that the theory is included in the above class, we choose a suitable background metric $f_{\mu\nu}$ that better accommodates a cosmological solution. Motivated by the Hassan and Rosen approach, we consider a massive cosmological solution that is continuous in the parameter space, in the sense that it is a solution of both massive and bimetric gravity theories in the limit of a vanishing kinetic term. In this respect, there is a particular choice for the background metric that relates it to the physical metric by a position-independent rescaling, namely, a positive constant $D^{2}$ (such as  in Refs. \cite{Roany2011,Baccetti2013})
\begin{equation}\label{eq:back_metric}
f_{\mu\nu} = D^{2}~g_{\mu\nu}.
\end{equation}

This is a well motivated choice for both, the Visser theory and for the Hassan and Rosen theory, since it leads to a massive tensor that acts effectively as a cosmological term in the Einstein field equations \cite{Roany2011,Baccetti2013}. Thereupon, it is straightforward to show that in both cases it is possible to write the $H_{\mu\nu}$ tensor as
\begin{equation}
H_{\mu\nu} = -\rho_{\rm eff} ~g_{\mu\nu},
\end{equation}
where here $\rho_{\rm eff}$ is an effective constant energy density coming from the massive term.

We can also define an effective pressure $p_{\rm eff}$ with equation of state $p_{\rm eff} = -\rho_{\rm eff}$, such that this term works like a cosmological constant. The specific dependence of $\rho_{\rm eff}$ on the parameters of each theory will be shown below. Before proceeding, let us write the general form of the Friedmann's equations and of the ECs for the two massive models. Considering that $g_{\mu\nu}$ is given by the Friedmann metric and that the energy-momentum tensor for the matter fields is $T_{\mu\nu} = (\rho + p)U_\mu U_\nu + pg_{\mu\nu}$, we find the Friedmann's equations
\begin{equation}\label{Friedmann1}
\left(\frac{\dot{a}}{a}\right)^2 = \frac{8\pi G}{3}(\rho + \rho_{\rm eff}) - \frac{k}{a^2},
\end{equation}
and
\begin{equation}\label{Friedmann2}
\frac{\ddot{a}}{a} + \frac{1}{2} \left(\frac{\dot{a}}{a}\right)^2 = - 4\pi G (p + p_{\rm eff}) - \frac{k}{2a^2},
\end{equation}
where $a(t)$ is the scale factor,  the dot corresponds to the derivative with respect to the cosmic time $t$,  and $k$ is the curvature of the spatial section. In what follows we use the definitions for the Hubble function $H \equiv H_0 E(a) \equiv \dot{a}/a$, the deceleration function $q = -\ddot{a}a/{\dot{a}^2}$ and the present value of the density parameter associated with the cosmological constant like term, namely  $\Omega_{\rm eff}^0 \equiv 8\pi G\rho_{\rm eff}/3H_0^2$. Hereafter the subscript and superscript 0 stand for the present-day quantities.

Therefore, by using Eqs.~\eqref{Friedmann1} and \eqref{Friedmann2} together with the results of Sec.~\ref{sec IIIA}, it is straightforward to show that SEC and DEC put, respectively, lower and upper bounds on $\Omega_{\rm eff}^0$, namely
\begin{align}
&\Omega_{\rm eff}^0 \geq \Omega_{\rm eff}^\mathrm{low}(z) \equiv -q(z) E^2(z), \label{eq:SEC} \\
&\Omega_{\rm eff}^0 \leq \Omega_{\rm eff}^\mathrm{up}(z) \equiv \frac{\left\{\left[2 - q(z)\right]E(z)^2 - 2\Omega_k^0(1+z)^2 \right\}}{3}, \label{eq:DEC}
\end{align}
where $1+z \equiv a_0/a$ and $\Omega_k^0 = -k/(a_0H_0)^2$ is the curvature density. Note that Eqs.~\eqref{eq:SEC} and \eqref{eq:DEC} refers respectively to
\begin{align}
& {\rm {\bf SEC}} & \rho + 3p \geq 0, \\
& {\rm {\bf DEC}} & \rho - p \geq 0.
\end{align}

Therefore, if we are able to determine from observations the curvature and the behavior of $q(z)$ and $E(z)$, it is possible to specify the kinetic constraints imposed by ECs on $\Omega_{\rm eff}^0$ and, consequently, on the parameters involved in each massive theory.  Moreover, the determination of these functions provide a whole range of upper and lower bounds, i.e., the inequalities~\eqref{eq:SEC} and~\eqref{eq:DEC} must be fulfilled in the entire range of redshift $z$ where they were determined. Accordingly, since we are dealing with inequalities we need only the maximum of $\Omega_{\rm eff}^\mathrm{low}(z)$ and the minimum of $\Omega_{\rm eff}^\mathrm{up}(z)$, that is
\begin{align}
z_1 &= \mathrm{arg \; max}\; \Omega_{\rm eff}^\mathrm{low},&  \Omega_{\rm eff}^\mathrm{low\star} &= \Omega_{\rm eff}^\mathrm{low}(z_1), \\
z_2 &= \mathrm{arg \; min}\; \Omega_{\rm eff}^\mathrm{up},&  \Omega_{\rm eff}^\mathrm{up\star} &= \Omega_{\rm eff}^\mathrm{up}(z_2).
\end{align}	

It is worth noting that for the same reconstruction $z_1$ and $z_2$ can be different. There is also the possibility that $\Omega_{\rm eff}^\mathrm{low\star} > \Omega_{\rm eff}^\mathrm{up\star}$, which means that the cosmological constant would be ruled out since it could not satisfy the bounds, indicating that the modification term should be at least time dependent.
In other words, the presence of lower and upper bounds, distributed in a wide range of redshifts, raises the possibility that the reconstructed curves lead to the largest lower bound (at $z_1$) larger than the smallest upper bound (at $z_2$). This is clearly inconsistent with an effective constant modification, we would need a time dependent $\Omega_\mathrm{eff}$ in order to satisfy both bounds. We explore this fact in Sec.~\ref{sec:results}.

\subsection{Density parameter for the Visser's Lagrangian}

Notice that the specific expressions for $\rho_{\rm eff}$ depend on the particular Lagrangian adopted. Recall that $\rho_{\rm eff} \equiv T^{\mu\nu}_{mass}U_{\mu}U_{\nu}$, where $U_\mu$ is the four velocity and $T^{\mu\nu}_{mass}$ is calculated by varying the Lagrangian [in particular, Eqs.~\eqref{eq:Visser_Lagrangian} and \eqref{eq:Hassan_Lagrangian}] with respect to $g_{\mu\nu}$ and considering that $f_{\mu\nu}$ and $g_{\mu\nu}$ are independent. For the Visser's Lagrangian \eqref{eq:Visser_Lagrangian} we find that $\rho_{\rm eff}$ is given by \cite{Roany2011}
\begin{equation}\label{eq:theory_1}
\rho_{\rm eff} = \frac{m^2}{16\pi G}(D^2 - 1).
\end{equation}

In principle, $\rho_{\rm eff}$ can be positive or negative, but it needs to be positive by requiring that $(D^2 - 1) > 0$ in order to be consistent with the recent phase of accelerated expansion of the Universe without the addition of any other component. The corresponding dimensionless density parameter is
\begin{equation}\label{omega theory 1}
\Omega_{\rm eff}^0 = \frac{\overline{m}^2}{6}(D^2 -1),
\end{equation}
where $\overline{m} \equiv m/H_0$.

In physical units the above mass parameter reads $\overline{m} = m/(\hbar H_0/c^2) = \ell_H/\lambda_g$, where $\ell_H = c/H_0$ is the Hubble distance and $\lambda_g = \hbar/m c $ is the Compton wavelength of the graviton. If $\lambda_g < \ell_H$,  i.e., $\overline{m} > 1$, this would mean that $m > \hbar H_0/c^2$ or $m > 2.13 \times 10^{-33}\,{\rm h}\,{\rm eV}/{\rm c}^2$ for $H_0 = 100\,{\rm h}\,{\rm km}\,{\rm s}^{-1}{\rm Mpc}^{-1}$.

\subsection{Density parameter for the Hassan and Rosen Lagrangian}

In the Hassan and Rosen approach, $\rho_{\rm eff}$ is related to the parameters $c_3$ and $c_4$, besides the mass $m$, as follows \cite{Baccetti2013}
\begin{equation}\label{theory 2}
\rho_{\rm eff} = \frac{3m^2}{8\pi G}Q(c_3,c_4),
\end{equation}
where
\begin{equation}\label{Q function}
Q(c_3,c_4) = \frac{1}{3}(1-D)^2[c_3(1-D) - 3],
\end{equation}
and $D$ is now given by
\begin{equation} \label{D scale}
D = 1 + \frac{3c_3}{2c_4} \pm \sqrt{\left( 1 + \frac{3c_3}{2c_4} \right)^2 - 1}.
\end{equation}

Note that $c_3(1-D) > 3$, in order to have $\rho_{\rm eff} > 0$. In this case, the corresponding density parameter reads
\begin{equation}\label{omegatheory2}
\Omega_{\rm eff}^0 = \overline{m}^2 Q(c_3,c_4).
\end{equation}

In the linear regime, the Visser's Lagrangian is clearly not free of ghosts for $f_{\mu\nu} = \eta_{\mu\nu}$. In this case the Visser's Lagrangian~\eqref{eq:Visser_Lagrangian} does not reduces to the FP Lagrangian, which is the only ghost free Lorentz invariant linear theory. On the other hand, it has been proved that the Hassan and Rosen Lagrangian \eqref{eq:Hassan_Lagrangian} is ghost-free at the complete non-linear level.

The massive gravity in the Hassan and Rosen approach may be viewed as a limit of bigravity for which there is a kinetic term for the background metric in addition to the interaction Lagrangian \eqref{eq:Hassan_Lagrangian}. However, as detailed discussed in \cite{Baccetti2013}, the limiting procedure is delicate. The authors found that solutions of bimetric gravity in the limit of vanishing kinetic term are also solutions of massive gravity, whereas the converse statement is not necessarily true. With this in mind and following \cite{Baccetti2013}, in the present article we consider a massive cosmological solution that is continuous in the parameter space, i.e., it is simultaneously a solution of both massive gravity and bimetric gravity in the limit of a vanishing kinetic term. 

Furthermore, it is worth mentioning that, besides the bimetric approach, there are also other methods to model a massive gravity. For instance, the introduction of an auxiliary extra dimension \cite{Gabadadze2009, Rham2010a} and the generation of mass through a gravitational Higgs mechanism \cite{Green1991, Siegel1994, Hooft2007, Kakushadze2008, Chamseddine2010}. These two alternatives can be formulated in terms of the bimetric massive gravity as discussed in \cite{Hassan2012} and \cite{Hassan2011}.

\section{Results}
\label{sec:results}

\begin{figure*}
\includegraphics[scale=1]{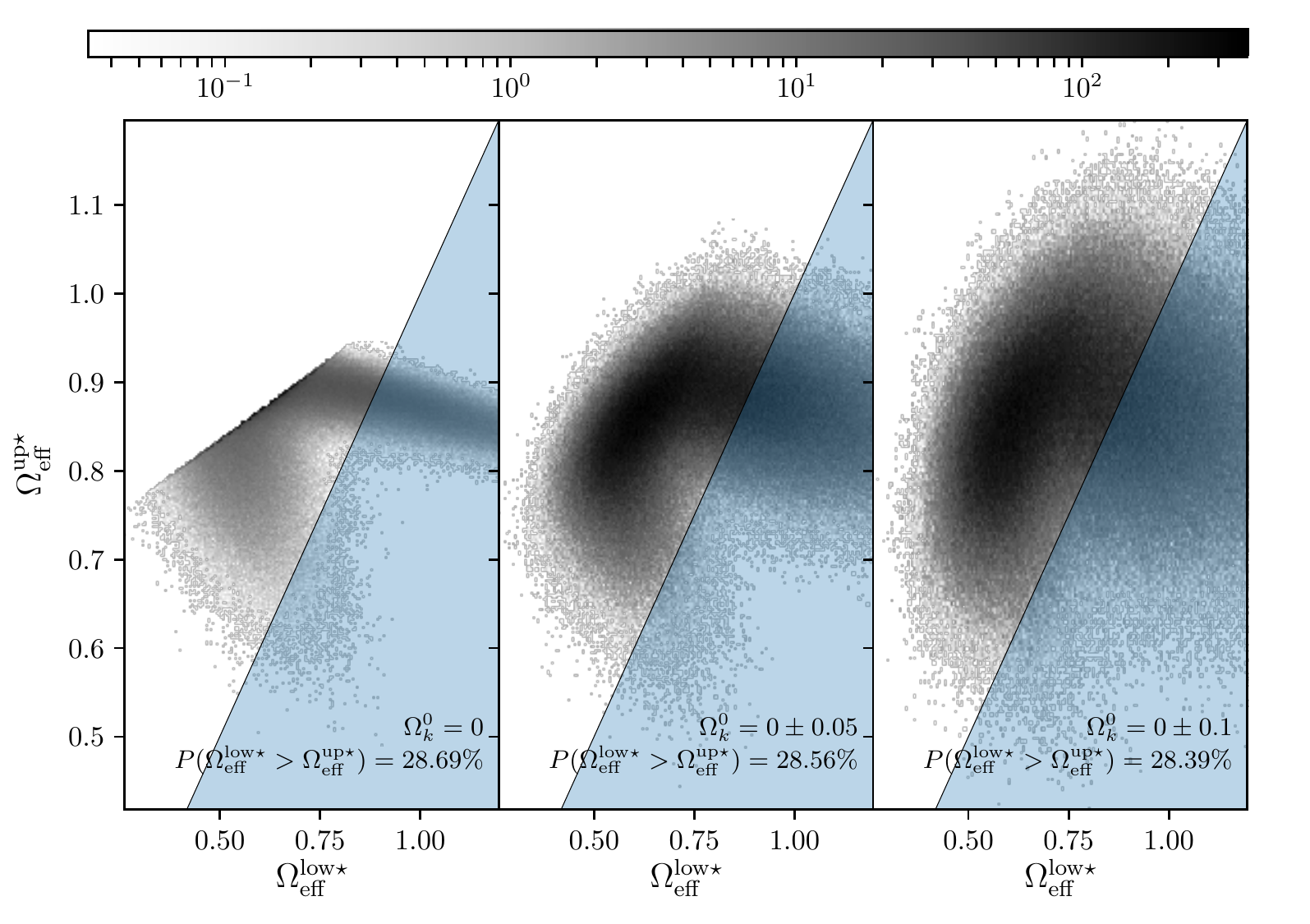}
\caption{\label{fig:posterior} The $(\Omega^\mathrm{low\star}_\mathrm{eff}, \Omega^\mathrm{up\star}_\mathrm{eff})$ posterior probability distribution for the three cases studied, i.e., flat universe (left panel) and considering Gaussian priors on $\Omega_k$ with zero mean and scatters equal to $0.05$ (middle panel) and $0.1$ (right panel). The shaded blue areas represents the regions where $\Omega_{\rm eff}^\mathrm{low\star} > \Omega_{\rm eff}^\mathrm{up\star}$. That is, a solution like the cosmological constant is rule out at this area of the parameter space.}
\end{figure*}

In Ref.~\cite{Penna-Lima2017} (the companion paper) we presented observational bounds for the energy conditions (considering both general relativity and extended theories of gravity) obtained from SNe Ia, BAO and $H(z)$ data. In particular, we used the JLA catalog of 740 SNe Ia  \cite{Betoule2014}, 11 BAO measurements \cite{Beutler2011, Font-Ribera2014, Ross2015, Delubac2015, Alam2016, Ata2017}, and 22 $H(z)$ data points \cite{Stern2010, Riess2011, Riess2011a, Moresco2012}. These data sets are comprised in the redshift interval $z \in [0, 2.33]$. We then estimated $q(z)$ and $E(z)$ in this redshift range, using the model-independent reconstruction method introduced by \citet{Vitenti2015a}. We applied the Markov Chain Monte Carlo (MCMC) approach, where we run about $5 \times 10^6$ points distributed among 50 chains for three different cases. These correspond to flat universe, $\Omega_k^0 = 0$, and two conservative Gaussian priors where $\Omega_k^0 = 0 \pm 0.05$ and $\Omega_k^0 = 0 \pm 0.1$.  For this, we made use of the MCMC ensemble sampler algorithm from the \textsf{NumCosmo} library (\textsf{NcmFitESMCMC}) \cite{Vitenti2014b} based on Ref.~\cite{Goodman2010}. The respective SNe Ia, BAO and $H(z)$ likelihoods are also implemented in \textsf{NumCosmo}. For details of these likelihoods and the data sets, see \cite{Penna-Lima2017}.

From these reconstructions of $q(z)$ and $H(z)$, in this work we obtain observational constraints for the upper and lower bounds for $\Omega_{\rm eff}^0$, see Eqs.~\eqref{eq:SEC} (SEC) and \eqref{eq:DEC} (DEC). For each point $p$ of the MCMC catalog, we calculated the functions $\Omega_{\mathrm{eff},p}^\mathrm{low}(z)$ and $\Omega_{\mathrm{eff},p}^\mathrm{up}(z)$ determining their maxima $(\Omega_{\mathrm{eff},p}^\mathrm{low\star})$ and minima $(\Omega_{\mathrm{eff},p}^\mathrm{up\star})$. These extremes were calculated considering different intervals 
$$
z \in [0, 0.5],\qquad z \in [0, 1.25], \qquad z \in [0, 1.5].
$$ 
The results concerning the upper and lower bounds of $\Omega_{\rm eff}^0$ are virtually independent of the choice of the intervals above. Nevertheless, using a larger redshift range increases the variances of $\Omega_{\mathrm{eff},p}^\mathrm{low\star}$ and $\Omega_{\mathrm{eff},p}^\mathrm{up\star}$, since the reconstructed curves have wider variance for higher $z$ as showed in \citet{Penna-Lima2017}. In practice this results in a larger probability of finding $\Omega_{\mathrm{eff},p}^\mathrm{low\star} > \Omega_{\mathrm{eff},p}^\mathrm{up\star}$}, which highlights the following trade-off. Our method constrains $\Omega_{\rm eff}^0$ in the redshift interval used to determine the extremes. Thus, more can be said about the behavior of $\Omega_{\rm eff}^0$ with a larger range, notwithstanding, as the interval increases so the variance. Here we carry out the analysis using the interval $z \in (0, 0.5)$. We emphasize that this choice makes almost no difference in the determination of $\Omega_{\mathrm{eff},p}^\mathrm{low\star}$ and $\Omega_{\mathrm{eff},p}^\mathrm{up\star}$. 

\begin{figure*}
\includegraphics[scale=1]{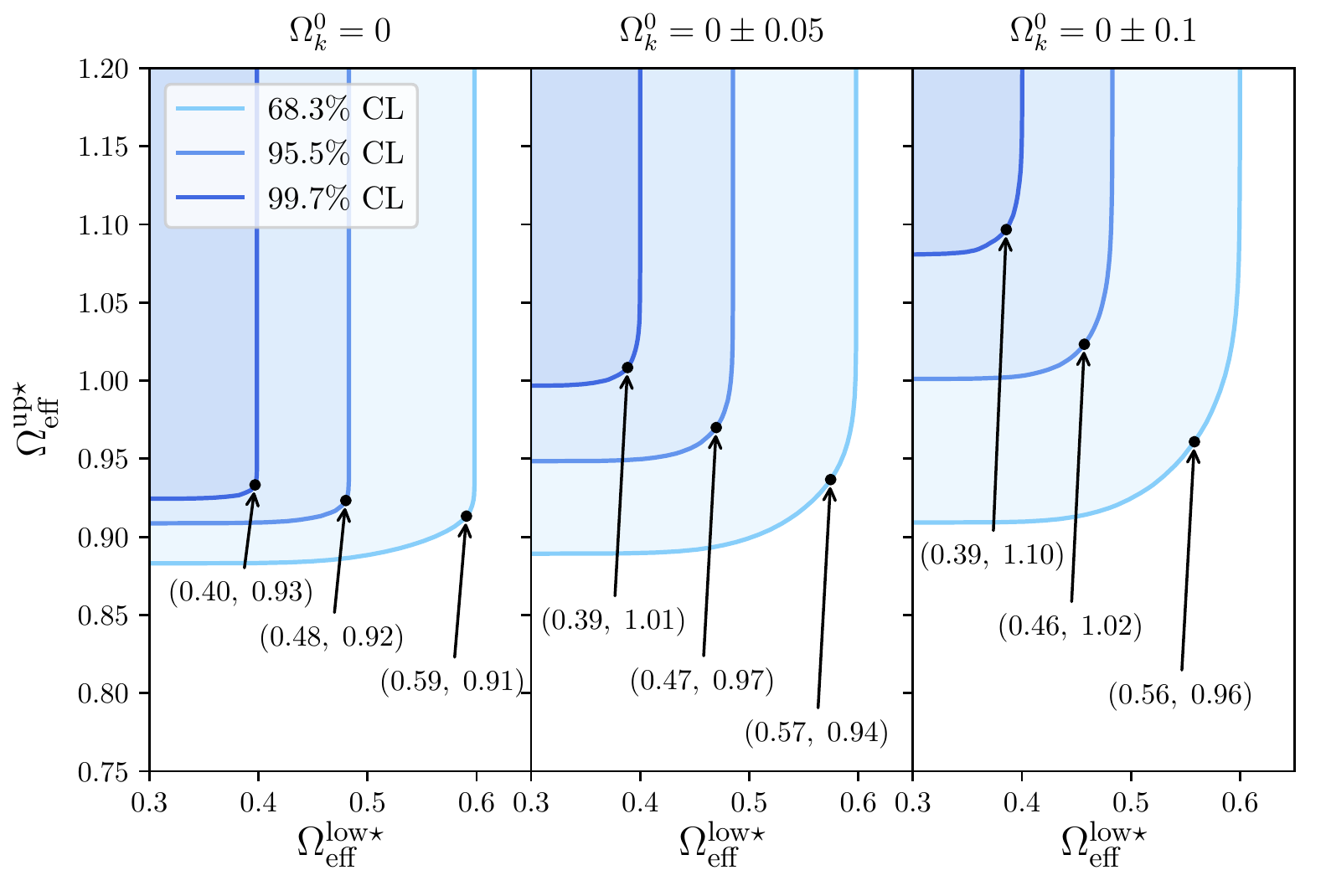}
\caption{\label{fig:omega_eff} The $1\sigma - 3\sigma$ contours of the $(\Omega^\mathrm{low\star}_\mathrm{eff}, \Omega^\mathrm{up\star}_\mathrm{eff})$ posterior probability distribution for the three cases studied, i.e., flat universe (left panel) and considering Gaussian priors on $\Omega_k$ with zero mean and scatters equal to $0.05$ (middle panel) and $0.1$ (right panel). We identify the points in the curves where we have the smallest intervals for a given confidence level. }
\end{figure*}

From our sample $(\Omega_{\mathrm{eff},p}^\mathrm{low\star}, \; \Omega_{\mathrm{eff},p}^\mathrm{up\star})$ we estimate the posterior probability density $P(\Omega_{\mathrm{eff}}^\mathrm{low\star}, \; \Omega_{\mathrm{eff}}^\mathrm{up\star})$ as showed in Fig~\ref{fig:posterior}. The probability distribution correlates both bounds in all three cases, and there are probabilities of about 30\% of finding $\Omega_{\mathrm{eff}}^\mathrm{low\star} > \Omega_{\mathrm{eff}}^\mathrm{up\star}$ (shaded blue areas). Note that this percentage is much more influenced by our choice of redshift interval discussed above. Larger intervals leads to regions where the curves are less constrained and consequently fluctuates and produces more points in the $\Omega_{\mathrm{eff}}^\mathrm{low\star} > \Omega_{\mathrm{eff}}^\mathrm{up\star}$ region.\footnote{For the other intervals, $z\in (0,1.25)$ and $z\in (0,1.5)$ we have, respectively, the probability of 48\% and 65\% of finding $\Omega_{\mathrm{eff}}^\mathrm{low\star} > \Omega_{\mathrm{eff}}^\mathrm{up\star}$.} In short, for all three cases, the cosmological constant like model has a 30\% probability of being  rejected. These cases suggest that a time dependent modeling would be necessary. Despite the probability be low and the cosmological constant be still allowed, this amounts to show that this method is capable of excluding a cosmological constant as the driver of the accelerated expansion in a model independent way.

In Fig.~\ref{fig:omega_eff} we show the $1\sigma -3\sigma$ confidence regions of $(\Omega^{\rm low\star}_{\rm eff}, \Omega^{\rm up\star}_{\rm eff})$. These contours were computed, respectively, as $P(\Omega^{\rm low\star}_{\rm eff} \geq x, \; \Omega^{\rm up\star}_{\rm eff} \leq y) = 68.27\%,$ 95.45\% and 99.73\%, where
\begin{equation}
\begin{split}
&P(\Omega^{\rm low\star}_{\rm eff} \geq x, \; \Omega^{\rm up\star}_{\rm eff} \leq y) = \\ &\int_0^y \mathrm{d}\Omega^{\rm up\star}_{\rm eff} \int_x^\infty \mathrm{d}\Omega^{\rm low\star}_{\rm eff} P(\Omega_{\mathrm{eff}}^\mathrm{low\star}, \; \Omega_{\mathrm{eff}}^\mathrm{up\star}). 
\end{split}
\end{equation}
That is, the contour curves correspond to intervals $(x_c,\; y_c)$, such that there is a probability of $31.73\%$, $4.55\%$ and $0.27\%$ (respectively $1\sigma$, $2\sigma$ and $3\sigma$) of finding $\Omega_{\mathrm{eff}}^\mathrm{low\star} < x_c$ and $\Omega_{\mathrm{eff}}^\mathrm{up\star} > y_c$. For a given $n\sigma$ contour curve the point $(0,\; y_c)$ is equivalent to marginalize the distribution in $\Omega_{\mathrm{eff}}^\mathrm{low\star}$ obtaining the point where the marginal probability of finding $\Omega_{\mathrm{eff}}^\mathrm{up\star}$ smaller than $y_c$ is $n\sigma$. Analogously, the point $(x_c,\; \infty)$, corresponds to the point where the marginal probability of finding $\Omega_{\mathrm{eff}}^\mathrm{low\star}$ larger than $x_c$ is $n\sigma$.

In particular, for the flat case, the smallest allowed intervals within  $1\sigma$, $2\sigma$ and $3\sigma$ confidence levels (CL) are, respectively,
\begin{align}
0.59 \leq \Omega_{\rm eff}^0\leq 0.91, \label{1S} \\
0.48 \leq \Omega_{\rm eff}^0\leq 0.92, \label{2S} \\
0.40 \leq \Omega_{\rm eff}^0\leq 0.93, \label{3S}
\end{align}
as one can see in Fig.~\ref{fig:omega_eff}. It is also worth noting that as we move in the $\Omega_{\mathrm{eff}}^\mathrm{low\star} \to 0$ direction the upper bound moves to the one obtained marginalizing over $\Omega_{\mathrm{eff}}^\mathrm{low\star}$ as discussed above, with the analogous happening when $\Omega_{\mathrm{eff}}^\mathrm{up\star} \to \infty$. From this we realize that the round corners of the contours result from the correlation between $\Omega_{\mathrm{eff}}^\mathrm{low\star}$ and $\Omega_{\mathrm{eff}}^\mathrm{up\star}$. This effect is more pronounced when $\Omega_k^0 \neq 0$, where we can see that the results obtained using individual marginalization of $\Omega_{\mathrm{eff}}^\mathrm{low\star}$ and $\Omega_{\mathrm{eff}}^\mathrm{up\star}$ would have lead to tighter but wrong constraints.

Since we are constraining the parameter $\Omega^0_{\rm eff}$, it is not possible to obtain tight constraints on each specific parameter of the massive gravity theories. However, it is possible to trace some conclusions about the parameter space as follows.

\begin{table}[t]
\centering
\caption{We show the upper bounds on the graviton mass mentioned in the main text and the corresponding lower bounds on $\log(D^2 - 1)$ for the Visser theory. To calculate the latter we considered $h=0.73$ and the lowest value of $\Omega_{\rm eff}^0$ in the interval of 3$\sigma$ CL.}
\label{graviton mass bounds}
\begin{tabular}{|c|c|c|}
\hline
$m_{\rm up}${[}eV/c$^2${]} & $\log(\bar{m}_{\rm up})$ & $[\log(D^2 - 1)]_{\rm low}$ \\ \hline \hline
$7.6 \times 10^{-20}$ \cite{Finn2002}  & 13.7                 & -27.0                  \\ \hline
$7.7 \times 10^{-23}$ \cite{Abbott2017} & 10.7                 & -21.0                  \\ \hline
$5.6 \times 10^{-28}$ \cite{Brandao2010, Brandao2010a} & 5.6                  & -10.8                  \\ \hline
$6 \times 10^{-32}$ \cite{Choudhury2004}    & 1.6                  & -2.82                  \\ \hline
\end{tabular}
\\
\end{table}

\begin{figure}
\centering
\includegraphics[width=1\linewidth]{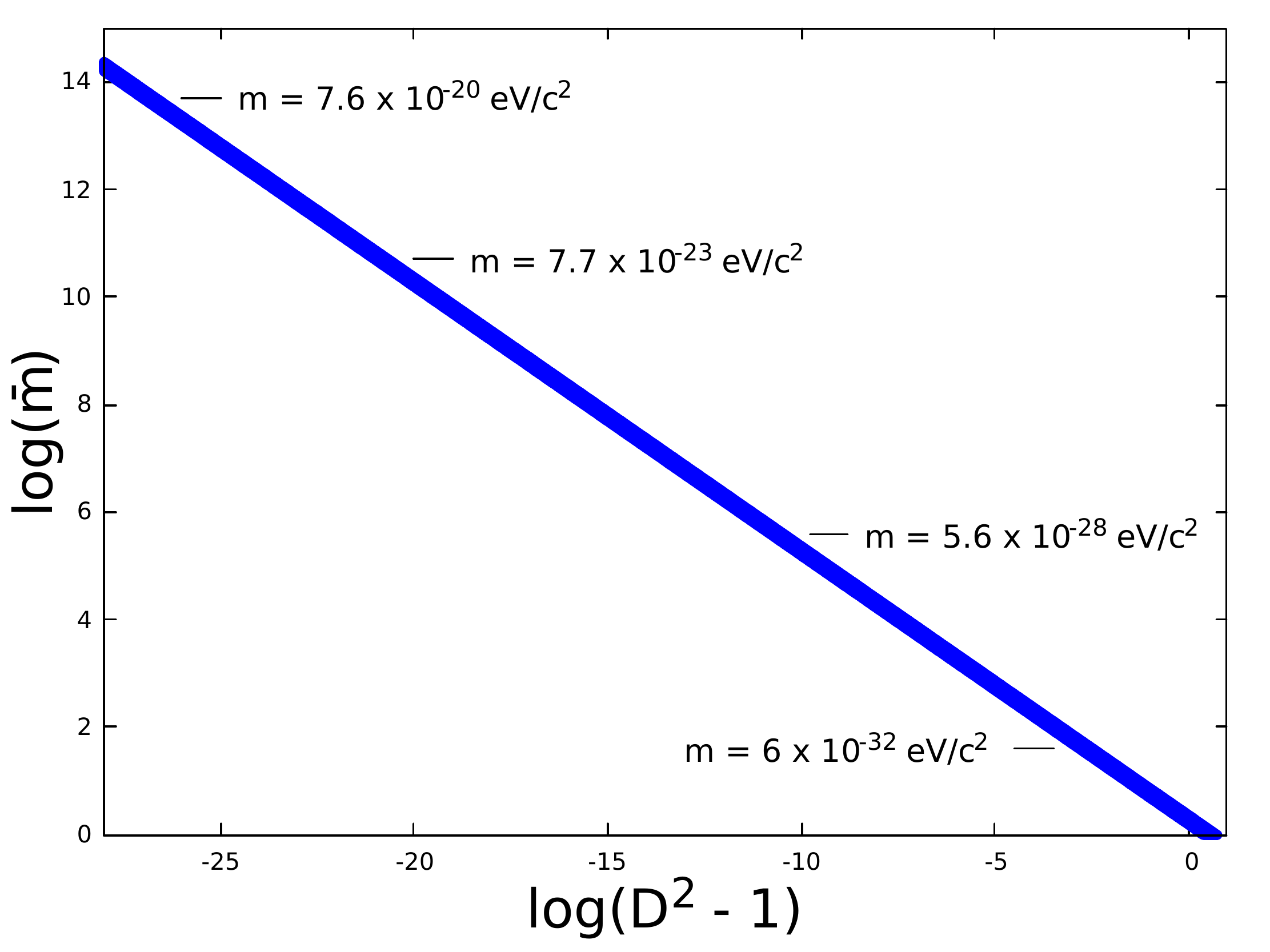}
\caption{\label{Fig1} We show the space parameter $\log(\overline{m})$ vs. $\log(D^2-1)$ for the Visser's theory considering $\Omega_{\rm eff}$ in the interval of 3$\sigma$ CL. The upper bounds on the graviton mass summarized in the Table \ref{graviton mass bounds} are also indicated.}
\end{figure}

For the case of the Visser's theory, the bounds on $\Omega_{\rm eff}^0$ implies bounds on the parameters $\overline{m}$ and $D$. In Fig.~\ref{Fig1} we show the constraints on the parameter space of the Visser's theory. The blue region is the $3\sigma$ CL of $\Omega_{\rm eff}^0$. It can be seen that the closer to one $D^2$ is, the higher is the graviton mass parameter. Since $f_{\mu\nu} = D^2 g_{\mu\nu}$, Fig.~\ref{Fig1} suggests that these metrics would differ only slightly from each other. 

In the non relativistic regime massive gravity reduces to a Yukawa-like potential instead of the Newtonian potential. Therefore, this can in principle be used to constraint $m$. This was already accomplished in Refs.~\cite{Araujo2007, Brandao2010, Brandao2010a}. In Ref.~\cite{Araujo2007} the authors consider analytical models of disk of spiral galaxies to constraint Yukawian potentials. They conclude that for disk galaxies exist $m_{g} < 5.6 \times 10^{-27}\; \mathrm{eV}/\mathrm{c}^2$. In addition, in Refs.~\cite{Brandao2010, Brandao2010a}, the authors use numerical simulations of spiral and elliptical galaxies to probe Yukawian potential. They find that consistent structures of galaxies can be obtained if $m_{g} < 5.6 \times 10^{-28}\;\mathrm{eV}/\mathrm{c}^2$. The aforementioned results imply the following constraint $\log(D^2 -1) > -10.8$. If we adopt such a lower bound as a reference we are lead to $D^2 \sim 1$ in the Visser's theory, which indicates again that $f_{\mu\nu}$ and $g_{\mu\nu}$ would differ slightly from each other.

There are other bounds on the graviton mass in the literature obtained by using a gravitational Yukawa potential. Two examples are the bound from the precession of Mercury which gives $m < 7.2\;\times 10^{-23}\; \mathrm{eV}/\mathrm{c}^2$~\cite{Talmadge1988, Will1998} and a stronger bound from weak lensing data of a cluster of stars at $z = 1.2$ is $m < 6 \times 10^{-32}\; \mathrm{eV}/\mathrm{c}^2$~\cite{Choudhury2004}. Where the last has the caveat of being dependent of the dark matter distribution and cosmological model.

In the dynamical regime, bounds on the graviton mass have been recently obtained using a modified dispersion relation in the observation of gravitational waves from the merger of binary black hole systems by the LIGO interferometer. In its third detection, an upper limit of $7.7 \times 10^{-23}\; \mathrm{eV}/\mathrm{c}^2$~\cite{Abbott2016a, Abbott2017} was established. Before LIGO, the bound due gravitational wave emission was $m < 7.6 \times 10^{-20}\; \mathrm{eV}/\mathrm{c}^2$~\cite{Finn2002} which was obtained from pulsar timing of PSR B1913+16 and PSR B1534+12. For a detailed review on the current and projected bounds on the graviton mass and the theoretical aspects related to such bounds see Ref.~\cite{Rham2017}. In Table \ref{graviton mass bounds} we summarize some upper bounds on the graviton mass and the corresponding lower bounds on $D$. 

For the case of the Hassan and Rosen theory, there are now three parameters, namely, $\overline{m}$, $c_3$ and $c_4$. In what follows we consider only the case with the plus sign before the square root in the expression of $D$ given by Eq.~\eqref{D scale}. Similar conclusions could be drawn by using the minus sign.

\begin{figure}
\centering
\includegraphics[width=1\linewidth]{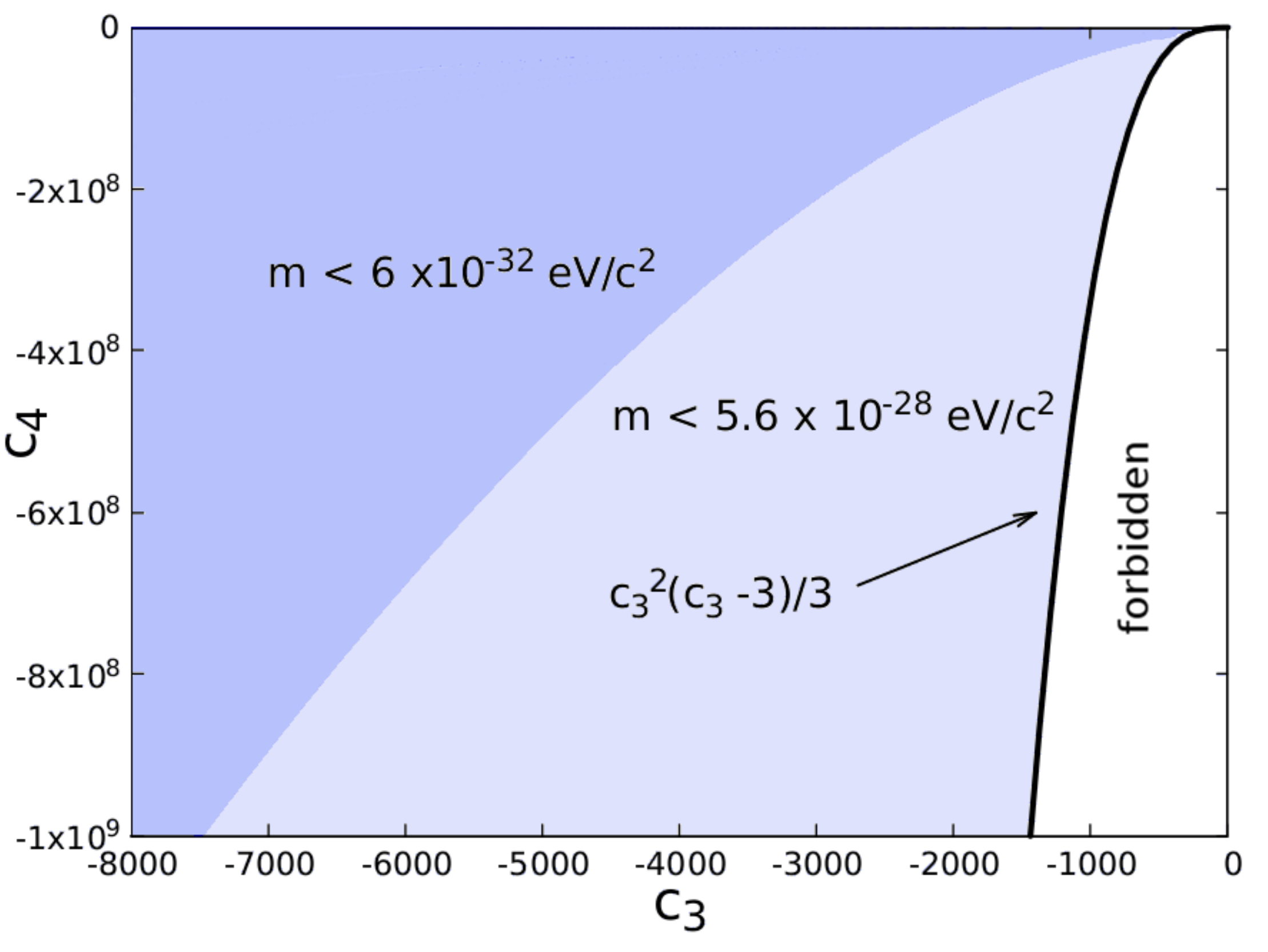}
\caption{The allowed region in the $c_4$ vs. $c_3$ plane for the Hassan and Rosen theory. The black line is the analytical function given by Eq. (\ref{c4 bounds}) which determines the maximum allowed region (obtained for $m_{\rm up} \rightarrow \infty$). The larger $m_{\rm up}$  is, the larger is the admissible space parameter  for $(c_4, c_3)$. As an example, we show the region for $m_{\rm up} = 6 \times 10^{-32}$ eV/c$^2$ and for $m_{\rm up} = 5.6 \times 10^{-28}$ eV/c$^2$ (which encompasses the region for $m_{\rm up} = 6 \times 10^{-32}$ eV/c$^2$). In the latter case, the region is almost coincident with the maximum allowed space parameter by virtue of the ranges used for $c_3$ and $c_4$. We used $h = 0.7$.}\label{Fig4}
\end{figure}

First of all, notice that $\Omega_\mathrm{eff}^0 = \Omega_\mathrm{eff}^0(\overline{m},c_3,c_4)$ given by Eq.~\eqref{omegatheory2} does not have real values for $c_4 > 0$, whereas $c_3$ can be positive or negative. Figure~\ref{Fig4} shows the parameter space where $c_3 < 0$. Notice also that, in order to have $\Omega_\mathrm{eff}^0(\overline{m}, c_3, c_4) > 0$, there is a forbidden area in the parameter space, as indicated. Such a region is delimited by the function
\begin{equation}\label{c4 lower}
c^\star_4 = \frac{1}{3} c_3^2(c_3 - 3),
\end{equation}
in the $(c_3, c_4)$ plane. Therefore, the maximum region for admissible pairs of values of these two parameters is
\begin{equation}\label{c4 bounds}
c^\star_4 < c_4 < 0, \qquad \mathrm{if} \qquad c_3 < 0.
\end{equation}

If we combine the bounds on $\Omega^0_{\rm eff}$ with an upper bound on the graviton mass we conclude that
\begin{equation}\label{bound on Q}
Q(c_3,c_4) > \frac{\Omega_{\mathrm{eff}}^\mathrm{low\star}}{\overline{m}^2_{\rm up}},
\end{equation}
where $Q(c_3, c_4)$ is given by Eq. (\ref{Q function}). The above equation gives the possible values of $c_3$ and $c_4$ given  $\Omega_{\mathrm{eff}}^\mathrm{low\star}$ and $m_{\rm up}$. In the Fig. \ref{Fig4} we have used the $3\sigma$ CL of $\Omega_{\rm eff}^0$, $h = 0.73$ and two values of $m_{\rm up}$, namely, $6 \times 10^{-32}$ eV/c$^2$ and $5.6 \times 10^{-28}$ eV/c$^2$. For the latter value, the region almost coincide with the maximum allowed space parameter in the ranges chosen, which can be found by imposing $m_{\rm up} \rightarrow \infty$ in the Eq. (\ref{bound on Q}) leading again to Eqs. (\ref{c4 bounds}) and (\ref{c4 lower}).

Notice that there is a large range for which $\vert c_4\vert \gg \vert c_3\vert$, in this case we obtain $D \sim 1$ from Eq.~\eqref{D scale} indicating that this theory also includes the case for which the difference between the dynamical and the background metrics is very small.

\section{Conclusion}
\label{sec:conclusions}

In this article we considered an arbitrary alternative gravity theory whose extra term acts like a cosmological constant. Then, using the strong and dominant energy conditions and the assumption that ordinary matter fulfills these conditions, we derived lower and upper bounds for $\Omega_\mathrm{eff}^0$. Considering three different priors for the curvature parameter ($\Omega_k^0$), we applied the reconstruction method using a set of low redshift data \cite{Vitenti2015a,Penna-Lima2017}. The bounds $\Omega_\mathrm{eff}^\mathrm{low}(z)$ and $\Omega_\mathrm{eff}^\mathrm{up}(z)$ limit the value of $\Omega_\mathrm{eff}^0$ in the considered redshift interval. On the grounds that both curves may have different shapes there are a non-zero probability of excluding a constant $\Omega_\mathrm{eff}^0$ since there are cases where $\Omega_\mathrm{eff}^\mathrm{low}(z) > \Omega_\mathrm{eff}^\mathrm{up}(z^\prime)$. In particular for $z\in[0,0.5]$  this probability is about 30\% for all three cases. In order to study  more closely these bounds, we obtained the bi-dimensional probability distribution of the extremes $\Omega_\mathrm{eff}^\mathrm{low\star}$ and $\Omega_\mathrm{eff}^\mathrm{up\star}$, finding the smallest allowed bounds in $1\sigma-3\sigma$ as exemplified for the flat case in Eqs.~(\ref{1S}--\ref{3S}).

It is worth emphasizing that we have presented in this work a model-independent procedure to test the nature of the dark energy in cosmological models like $\Lambda$CDM, which in turn can stem from many possible modifications in the gravitational theory. Furthermore, by assuming that ordinary matter do not violate the EC's we were able to impose both lower and upper bounds on $\Omega_\mathrm{eff}^0$. The value of these constraints are in accordance with the model dependent analysis, which leads to $\Omega_\mathrm{eff}^0 = \Omega_\Lambda \approx 0.7$ (see for example~\cite{PlanckCollaboration2015}). This shows two independent analyses pointing out in the same direction. 

In light of these results we also studied the constraints for two bimetric massive gravity theories, namely, the Visser's and the Hassan and Rosen's. We have taken into account the particular case for which the massive term mimics a cosmological constant in the Einstein's field equations. Besides such a solution can potentially generate the present acceleration of the Universe, it is also motivated by the fact that it is continuous in the parameter space, in the sense that it is a solution simultaneously of massive gravity and bigravity in the limit of vanishing kinetic term.

By considering the SEC and DEC we also have imposed constraints in the parameter space of both massive theories. Particularly, in the context of the Hassan and Rosen's approach, we identified a forbidden region in the $(c_3,\; c_4)$ plane which is independent of the graviton mass. But, in general, it is not possible to obtain tight constraints on each specific parameter of the theories since the ECs involve essentially $\Omega_{\rm eff}^0$ which is a combination of the $c_3$, $c_4$ and $\overline{m}$ parameters. In this sense, we establish a wide range of possible values of $c_3$ and $c_4$ which are in accordance with the bounds for the graviton mass. For instance, considering the maximum allowed region in the space parameter,  we have $c_4^\star < c_4 < 0$ where the lower bound for $c_4$ is given by Eq. (\ref{c4 lower}). Therefore, if $c_3 = -1\rightarrow c_4^\star = -1.33$ while if $c_3 = -10^3 \rightarrow c_4^\star = -3.34 \times 10^8$. In order to impose further constraints in such a space parameter, other gravitational tests designed to bound $c_3$ and $c_4$ (independent of $\Omega_{\rm eff}$) would be needed.    

\begin{acknowledgments}
MESA and JCNA would like to thank the Brazilian agency FAPESP for financial support (grant 13/26258-4). JCNA thanks also the Brazilian agency CNPq (Grants 308983/2013-0; 307217/2016-7) by the financial support. FCC was supported by FAPERN/PRONEM and CNPq. MPL acknowledges Labex ENIGMASS and CNPq (PCI/MCTIC/CBPF program) for financial support. SDPV thanks BELSPO non-EU postdoctoral fellowship. This research was performed using the Mesu-UV supercomputer of the Pierre \& Marie Curie University | France (UPMC) and the computer cluster of the State University of the Rio Grande do Norte (UERN) | Brazil.
\end{acknowledgments}

\bibliographystyle{apsrev}
\bibliography{paper_ec_massive_arxiv}

\end{document}